\documentclass{ws-procs9x6}

\newcommand{\be}{\begin{equation}}
\newcommand{\ee}{\end{equation}}
\newcommand{\bea}{\begin{eqnarray}}
\newcommand{\eea}{\end{eqnarray}}
\newcommand{\bfk}{\mbox{\boldmath $k$}}

\newcommand{\pup}{p^\uparrow}

\begin{document}

\title{The gluon Sivers distribution \\ in $D$ production at RHIC 
%\pup \!\! \p \to D \, X$
}

\author{M. ANSELMINO AND {\underline{M. BOGLIONE}}}

\address{Dipartimento di Fisica Teorica, Universit\`a di Torino and \\
          INFN, Sezione di Torino, Via P. Giuria 1, I-10125 Torino, Italy}

\author{E. LEADER}

\address{Imperial College, Prince Consort Road, London, SW7 2BW, England}  

\author{U. D'ALESIO AND F. MURGIA}

\address{Dipartimento di Fisica, Universit\`a di Cagliari and \\
        INFN, Sezione di Cagliari, C.P. 170, I-09042 Monserrato (CA), Italy}

\maketitle

\abstracts{The single transverse spin asymmetry in $D$ meson production 
at RHIC can provide a clean measure of the gluon Sivers distribution function. 
At intermediate rapidity, $D$ production is largely 
dominated by the elementary $gg \to c \bar c$ channel, where there cannot be 
any transverse spin transfer. Therefore, any transverse single spin asymmetry 
observed for $D$'s produced in $\pup p$ interactions can only originate
from the Sivers effect in the gluon distribution functions.
A sizeable transverse single spin asymmetry measured by PHENIX or STAR 
experiments would then give direct information on the size of the gluon Sivers 
distribution function.}

Parton distribution and fragmentation functions are phenomenological 
quantities which have to be 
obtained from experimental observation and cannot be theoretically predicted. 
When parton intrinsic transverse momenta are taken into account, 
a large number of pdf's and ff's arise: the main difficulties in gathering 
experimental information on these spin and $\bfk_\perp$ dependent functions
is that most often more of them  contribute to the same physical observable, 
making it extremely hard to estimate each single one separately.
The Sivers function $\Delta^N f (x, \bfk _\perp)$ \cite{siv}, which describes  
the probability density of finding unpolarized partons inside a transversely 
polarized proton, is one of these functions. 
It plays a crucial role since it can explain single spin asymmetries in terms 
of parton dynamics \cite{noi1}.  

We suggest here an experiment to be conducted at RHIC which can isolate 
the gluon Sivers effect, making it possible to reach direct independent 
information on the gluon Sivers distribution function 
$\Delta^N f_{g/\pup} (x, \bfk_\perp)$. A full description can be found in a 
detailed recently published paper \cite{D-paper}.

\begin{figure}[t]
%\epsfxsize=5cm   %width of figure - will enlarge/reduce the figures
%\epsfbox{fig3.eps}
%\figurebox{2cm}{3cm}{} %to have a box alone 
\centerline{\epsfxsize=3.7in\epsfbox{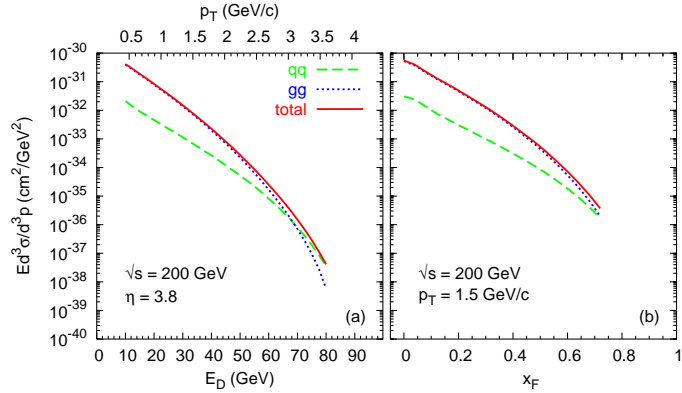}}   
\caption{ \label{unp-sigma}The unpolarized cross section for the process 
$pp \to DX$ at $\sqrt s = 200$ 
GeV, as a function of $E_D$ and $p_T$ at fixed pseudo-rapidity $\eta=3.8$ (a),
and as a function of $x_F$ at fixed transverse momentum $p_T=1.5$ GeV/$c$ (b).
}
\end{figure}

Let us consider the single spin asymmetry 
$A_N=(d\sigma ^\uparrow - d\sigma ^\downarrow )/
     (d\sigma ^\uparrow + d\sigma ^\downarrow )$ 
for $\pup p \to DX$ processes at RHIC energy, $\sqrt{s} = 200$ GeV. These 
$D$ mesons originate from $c$ or $\bar c$ quarks, which at LO can be created 
either via a $q \bar q$ annihilation, $q \bar q \to c \bar c$, or via a gluon 
fusion process, $gg \to c \bar c$. The elementary cross section for 
the fusion process is  much larger than the $q \bar q \to c \bar c$ 
cross section.
Therefore, the gluon fusion dominates the whole $\pup p \to DX$ process, 
up to $x_F \simeq 0.6$. 
In Fig.~\ref{unp-sigma}(a) we show the unpolarized cross section for the 
process $p\,p \to DX$ at $\sqrt s = 200$ GeV as a function of both the 
heavy meson energy $E_D$ and its transverse momentum $p_T$, at fixed 
pseudo-rapidity $\eta = 3.8$. 
In Fig.~\ref{unp-sigma}(b) the same total cross section is presented as 
a function of $x_F$ at fixed $p_T = 1.5$ GeV/$c$.    
The dashed and dotted lines correspond to the $q\bar q \to c\bar c$ and 
$gg \to c\bar c$ contributions respectively. 
These plots clearly show the striking dominance of the $gg \to c\bar c$ 
channel over most of the $E_D$ and $x_F$ ranges covered by RHIC kinematics. 
%%%%%%%%%%%%%%%%%%%%%%%%%%%%%%%%%%%%%%%%%%%%%%%%%%%%%%%%%%%%%%%%%%%%%%%%%%%
%%%%%%%%%%%%%%%%%%%%%%%%%%%%%%%%%%%%%%%%%%%%%%%%%%%%%%%%%%%%%%%%%%%%%%%%%%%

As the gluons cannot carry any transverse spin, the elementary process 
$gg \to c \bar c$ results in unpolarized final quarks. In the 
$q \bar q \to c \bar c$  process one of the initial partons can be polarized; 
however, at lowest order there is no 
single spin transfer in this $s$-channel interaction so that the final 
$c$ and $\bar c$ are again not polarized.  
Consequently, the charmed quarks fragmenting into the observed $D$ 
mesons cannot be polarized, and there cannot be any Collins 
fragmentation effect.      

Therefore, transverse single spin asymmetries in $\pup p \to DX$ can only be 
generated by the Sivers mechanism, namely a spin-$\bfk_\perp$ asymmetry in 
the distribution of the unpolarized quarks and gluons inside the polarized 
proton, coupled to the unpolarized interaction process 
$q \bar q \to c \bar c$ and dominantly $gg \to c \bar c$, and
the unpolarized fragmentation function of either the $c$ or the $\bar c$ 
quark into the final observed $D$ meson. Detailed formulae for the $A_N$ generated by the Sivers effect can be found in a recent paper by U. D'Alesio and F. Murgia \cite{unp}.

The Sivers distribution functions \cite{siv} 
for quarks and gluons are defined by 
\be
\Delta ^N f_{a/\pup}(x_a,\bfk_{\perp a}) = 
%\hat f_{a/\pup} (x_a,\bfk_{\perp a}) - \hat f_{a/\pdown}(x_a,\bfk_{\perp a})= 
\hat f_{a/\pup} (x_a,\bfk_{\perp a}) - \hat f_{a/\pup}(x_a,-\bfk_{\perp a})\,,
\label{siv}
\ee
where $a$ can either be a light quark or a gluon.

Here, we consider intrinsic 
transverse motions in the distributions of initial light quarks, in the 
elementary process and in the heavy quark fragmentation function, {\it i.e.} 
we consider a fully non planar configuration for the partonic scattering.  
This has two main consequences: on one side, taking into account three 
intrinsic transverse momenta makes the kinematics highly 
non trivial; on the other side it generates a large number of contributions, 
other than the Sivers effect, which originate from all possible combinations 
of $\bfk_\perp$ dependent distribution and fragmentation functions, each 
weighted by a phase factor. 
This is a crucial topic, extensively discussed by U. D'Alesio in these 
proceedings and in a paper dedicated to the suppression of Collins and other 
effects in proton-proton scattering \cite{noi3}.
We have explicitely verified that all contributions to the 
$\pup p \to DX$ single 
spin asymmetry from $\bfk_\perp$ dependent pdf's and ff's, aside from those 
of the Sivers functions, are multiplied by phase factors which make the 
integrals over the transverse momenta either negligibly small or 
identically zero.

In the previous literature, analyses of single spin asymmetry data were 
performed under the assumption $\Delta ^N f_{g/\pup}=0$. RHIC data on $A_N$ 
in $\pup p \to DX$ will enable us to test the validity of this assumption.
In fact, as the $gg\to c\bar c$ elementary scattering largely dominates the 
process up to $x_F\simeq 0.6$ (see Fig.~1), any sizeable single spin 
asymmetry measured in $\pup p \to DX$ at moderate $x_F$'s would be the direct 
consequence of a non zero contribution of $\Delta ^N f_{g/\pup}$. 

Since we have no information about the gluon Sivers function, we consider two 
opposite extreme scenarios: the first being the case in which 
the gluon Sivers function is set to zero, 
$\Delta ^N f_{g/\pup}(x_a,\bfk_{\perp a}) = 0$, and the quark Sivers function 
$\Delta ^N f_{q/\pup}(x_a,\bfk_{\perp a})$ is taken to be at its maximum 
allowed value at any $x_a$; the second given by the opposite situation, where 
$\Delta ^N f_{q/\pup}=0$ and $\Delta ^N f_{g/\pup}$ is maximized in $x_a$.

\begin{figure}[t]
%\epsfxsize=5cm   %width of figure - will enlarge/reduce the figures
%\epsfbox{fig3.eps}
%\figurebox{2cm}{3cm}{} %to have a box alone 
\centerline{\epsfxsize=3.7in\epsfbox{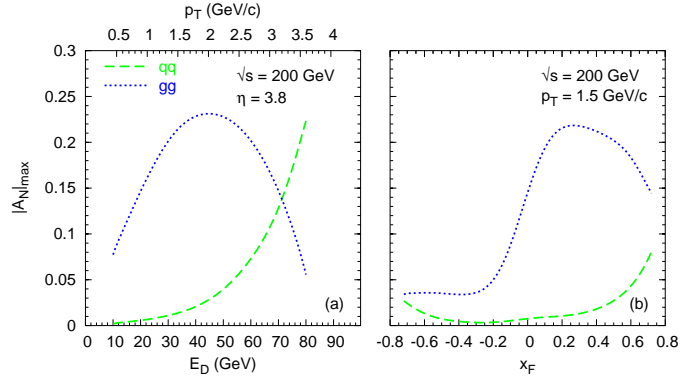}}   
\caption{\label{A-N} Maximized values of $|A_N|$ for the process
$\pup p \to DX$ as a function of
$E_D$ and $p_T$ at fixed pseudo-rapidity (a), and as a function of $x_F$ at
fixed transverse momentum (b). The dashed line corresponds to a
maximized quark Sivers function (with the gluon Sivers function set to zero),
while the dotted line corresponds to a maximized gluon Sivers function (with
the quark Sivers function set to zero).}
\end{figure}

Fig.~\ref{A-N} shows our estimates for the maximum value of the single spin 
asymmetry in $\pup p \to DX$. The dashed line shows $|A_N|$ when the quark 
Sivers function is set to its maximum, {\it i.e.} 
$\Delta ^N f _{q/\pup}(x) = 2f _{q/p}(x)$, 
while setting the gluon Sivers function to zero. 
Clearly, the quark contribution to $A_N$ is very small over most of the 
kinematic region, in both cases (a) and (b).
The dotted line corresponds to the SSA one finds in the opposite situation, 
when $\Delta ^N f _{g/\pup}(x) = 2f _{g/p}(x)$ and $\Delta ^N f _{q/\pup} =0$:
here the asymmetry presents a sizeable maximum in the central $E_D$ 
and positive $x_F$ energy region. This particular shape is given by the 
azimuthal dependence of the numerator of $A_N$~\cite{D-paper}. 

By looking at Fig.~\ref{A-N} it is natural to conclude that any sizeable 
single transverse spin asymmetry measured by STAR or PHENIX experiments
at RHIC in the region $E_D\le 60$ GeV or $-0.2 \le x_F \le 0.6$, would give 
direct information on the size and importance of the gluon Sivers function.  
%%%%%%%%%%%%%%%%%%%%%%%%%%%%%%%%%%%%%%%%%%%%%%%%%%%%%%%%%%%%%%%%%%%%%%%%%%%
%%%%%%%%%%%%%%%%%%%%%%%%%%%%%%%%%%%%%%%%%%%%%%%%%%%%%%%%%%%%%%%%%%%%%%%%%%%

\end{document}